\documentclass[12pt]{article}
\usepackage{graphicx}
\usepackage{dcolumn}
\usepackage{bm}

\def\Trs{\mathop{\rm tr}}         

\def\trs{{\rm tr_s}\,}
\def\smn{\sigma_{\mu\nu}\,}
\def\dov{D^{ov}}
\def\invsqYY{\frac{1}{\sqrt{Y^\dagger Y}}}
\def\invsig#1{\frac{1}{\sigma^2 + #1^\dagger #1}}
\def\invsigz{\frac{1}{\sigma^2 + z}}
\def\intdk{\int_{-\pi}^\pi\frac{d^4k}{(2\pi)^4}}
\def\intds{\int_{-\infty}^\infty d\sigma\,}
\def\da{\frac{{\rm d}}{{\rm da}}}
\def\dasq{\frac{{\rm d^2}}{{\rm da^2}}}
\def\one{{\bf 1}}

\def\sla#1{
\setbox0=\hbox{#1}
\setbox1=\hbox{\slash}
\dimen0 = \wd0
\advance \dimen0 by \wd1
\divide \dimen0 by 2
\box0\kern-\dimen0\box1
}

\def\sD{\sla{D}}
\def\stD{\tilde{\sD}}
\def\ss{\sla{s}}
\def\tR{\tilde{R}}
\def\tD{\tilde{D}}
\def\half{\frac{1}{2}}
\def\id{\mbox{1}}

\def\beq{\begin{equation}}
\def\eeq{\end{equation}}
\def\beqa{\begin{eqnarray*}}
\def\eeqa{\end{eqnarray*}}
\def\beqan{\begin{eqnarray}}
\def\eeqan{\end{eqnarray}}

\def\rb{\right)}
\def\lb{\left(}

\begin{document}


\begin{center}

{\bf {\LARGE Gauge Field Strength Tensor}}

\vspace{0.4cm}

{\bf {\LARGE from the Overlap
             Dirac Operator}}

\vspace{1.5cm}

{\bf K.F. Liu, A. Alexandru, and I. Horv\'{a}th}

{\it Dept.\ of Physics and Astronomy, University of Kentucky,
Lexington, KY 40506}

\end{center}

\begin{abstract}
We derive the classical continuum limit of the operator
tr$_s\, \sigma_{\mu\nu}D^{ov}(x,x)$ with $D^{ov}$ being the overlap 
Dirac operator and show that it corresponds to the gauge 
field strength tensor $F_{\mu\nu}(x)$.
\end{abstract}

\vfill
\newpage

\section{Introduction}

Lattice gauge operators are usually constructed explicitly from link
variables. For example, the Wilson gauge action uses the product 
of gauge links at the boundary of the square plaquette.
Similarly, the gauge field strength tensor can be defined
through suitable combinations of such elementary plaquettes. To
improve scaling behavior of the action and other gauge operators,
rectangular and more sophisticated loops have been
included~\cite{iw83,lw85,DBW2,hn94}. Furthermore, it was shown that
operators with smeared gauge links, being less ultralocal, are
effective in filtering out ultraviolet fluctuations and improving the
efficiency of numerical calculations~\cite{tep86,APE87,HYP01,mp04}.

A different approach is realized in constructing topological charge densities
via Ginsparg-Wilson Dirac operators~\cite{hln98}. 
It uses matrix elements of the Dirac operator as a starting point. By virtue 
of the fact that this type of Dirac operators are inevitably 
non-ultralocal~\cite{nonultr}, they entail a sum of all gauge 
loops and are thus automatically smeared. While technically involved, this 
approach has the advantage that Ginsparg-Wilson operators incorporate an
exact lattice chiral symmetry~\cite{lus98}. This leads to an index theorem on
the lattice~\cite{hln98}.

The overlap Dirac operator~\cite{neu98a} offers a concrete example 
which satisfies Ginsparg--Wilson relation~\cite{neu98b}. Its compact form 
makes it amenable to analytic as well as numerical calculations. The 
associated topological charge density was shown to have the correct 
classical continuum limit via weak-coupling expansion~\cite{ky99} and 
by direct calculations~\cite{ada02,fuj99,suz99}. 
It is with this topological density operator that the sub-dimensional 
long-range structure has been discovered~\cite{hdd03b,haz05a,ahz05} 
and confirmed~\cite{ikk06} in QCD, as well as in 2-D CP(N-1)
models~\cite{alt05}. It is also with this operator that the
required negativity of the topological density correlator is
borne out clearly with only a handful of configurations~\cite{haz05b}. 
Whereas, detecting this negativity to the same precision using conventional 
operators, such as those used in the glueball calculation~\cite{cad06}, would  
require much larger statistics. 
Based on these observations, it was suggested by one of the authors 
that the condition of chiral symmetry plays a relevant role 
in efficient suppression of the ultraviolet noise~\cite{hov06a}, 
and that all gauge operators can be constructed from the chirally 
symmetric Dirac operator~\cite{hov06b}. This way, one can also 
have a formulation of lattice QCD where the gauge action, 
the $\theta$ term and the fermion action are all expressed in 
terms of the lattice Dirac operator~\cite{hov06b}.

    In the present work, we will concentrate on the gauge field strength 
tensor. 
It was suggested that the classical limit of the 
tensor component of the overlap operator, $\trs \smn D^{ov}(x,x)$, 
is proportional to the gauge field strength tensor 
$F_{\mu\nu}(x)$~\cite{hov06b,Nie98A}. We shall explicitly calculate it and
show that this is indeed the case. In view of the fact the chirally symmetric 
Dirac operators are non--ultralocal~\cite{nonultr},
the derivation is somewhat non--trivial. 
Since the overlap Dirac operator is expected
to be local on gauge configurations of interest~\cite{ov_loc,draper}, our 
result implies that the quantum operator constructed this way represents a 
valid definition of gauge field strength tensor in lattice QCD. This could 
provide a practical tool for evaluating gluonic observables. A preliminary 
version of the calculation was given in Ref.~\cite{liu06}.

\section{Formulation}

Our goal in this paper is to show that if we
discretize a classical gauge field $A_\mu(x)$ on a hypercubic lattice
with (classical) lattice spacing $a$, and consider the overlap Dirac
operator $D^{ov}$ on such background, then
\begin{equation}
    {\Trs}_s\, \sigma_{\mu\nu} \, D^{ov}(x,x) \;\propto\;
    a^2\, F_{\mu\nu}(x)
    \qquad \mbox{\rm for} \qquad
    a\longrightarrow 0
    \label{eq:5}
\end{equation}
where ``$\Trs_s$'' denotes the spinor trace.~\footnote{Note that we
use the convention that the real--valued arguments of lattice
quantities (such as $D(x,x)$ on LHS of Eq.~(\ref{eq:5})) are given in
parenthesis, while the integer--valued lattice coordinates are written
as subscripts (such as $D_{n,n}$).}
By classical SU(3) gauge configuration we mean any configuration of the 
gauge field that is smooth (differentiable arbitrarily many times) almost 
everywhere.
In the above equation we have implicitly assumed that $x$ is not
a singular point of $A_\mu(x)$. It was also implicitly assumed that
point $x$ is a lattice grid point of a superimposed lattice for arbitrary
lattice spacing $a$. This is technically most easily (and without loss
of generality) achieved if we focus on the point $x=0$ with the origin
of the lattice coordinate system aligned with the Cartesian coordinate
system of $R^4$. Also, to avoid an extensive discussion of technical
issues associated with the transcription of the field with singularities 
onto the hypercubic lattice, we will implicitly assume in the following
that the field is smooth everywhere. This is sensible since due to
the locality of the overlap operator, the result is expected
to be valid for an arbitrary non-singular space--time point $x$. For a 
discussion relevant to this 
point, the reader is referred to Ref.~\cite{ada02}.

  In the convention that we will use, the continuum gauge
potential  $A_\mu(x)$ is the vector field of traceless Hermitian
matrices~\footnote{Note that this differs from conventions of
Ref.~\cite{hov06b}, where anti--Hermitian gauge potentials were used
instead. The equations below can be obtained from equations of
Ref.~\cite{hov06b} via substitutions $A_\mu(x) \rightarrow i A_\mu(x)$,
$F_{\mu\nu}(x) \rightarrow i F_{\mu\nu}(x)$. The value of constant
$c^T$ in Eq.(\ref{eq:70}) is the same in both conventions.} and 
the corresponding field--strength tensor is
\begin{equation}
      F_{\mu \nu}(x) \,\equiv\,
      \partial_\mu A_\nu(x) - \partial_\nu A_\mu(x) +
      i \,[\, A_\mu(x), A_\nu(x) \,].
   \label{eq:10}
\end{equation}
With the covariant derivative defined as
\begin{equation}
   D_\mu \,\phi(x) \,=\, (\partial_\mu + i A_\mu(x)) \,\phi(x),
\end{equation}
one has 
\begin{equation}   \label{eq:20}
   [D_\mu,D_\nu] \, \phi(x) \,=\, i F_{\mu \nu}(x) \, \phi(x).
\end{equation}

The transcription of $A_\mu(x)$ to the hypercubic lattice with
integer coordinates \mbox{$n \equiv (n_1,n_2,n_3,n_4)$} is accomplished
in a standard manner. If $a$ is the classical lattice spacing,
we associate the lattice site $n$ with the space--time point $x=an$,
and the lattice link variable $U_{n,\mu}$ is defined as
\begin{equation}
   U_{n,\mu}(a) \,\equiv\, \exp\Bigl(i a A_\mu(an)\Bigr)
   \label{eq:40}
\end{equation}

The overlap Dirac operator $\dov$ is given by~\cite{neu98a}
\beq
\dov \,=\, \rho \lb \id +X\frac{1}{\sqrt{X^\dagger X}}\rb; \qquad\quad
 X  = \sD - R - \rho + 4r
 \label{eq:50}
\eeq
where $-\rho$, $\rho\in (0,2r)$, is the negative mass parameter and
\beq  \label{DR}
 D_\mu = \half [U_\mu S_\mu - S_\mu^\dagger U_\mu^\dagger] \qquad\quad
 R \,=\, \frac{r}{2} \sum_\mu [U_\mu S_\mu + S_\mu^\dagger U_\mu^\dagger]
 \label{eq:60}
\eeq
with
\beq
(S_\mu)_{m,n} \,\equiv\, \delta_{m,n-\hat{\mu}} \qquad\quad
(U_\mu)_{m,n} \,\equiv\, U_{m,\mu}\,\delta_{m,n}
 \label{eq:65}
\eeq
We shall take the Euclidean $\gamma$--matrices
to be Hermitian, i.e. $\gamma_\mu^\dagger = \gamma_\mu$ and
$\{\gamma_\mu,\gamma_\nu\}=2\delta_{\mu,\nu}$.

With the above defining relations, we shall proceed to show the following 
in an explicit calculation:

\noindent {If $A_\mu(x)$ is a smooth SU(3) gauge potential
on $R^4$, and $U(a)$ is the transcription of this field to the hypercubic
lattice with classical lattice spacing $a$, then
\begin{equation}   \label{ct}
   {\Trs}_s\, \sigma_{\mu\nu} \, \dov_{0,0}\Bigl(U(a)\Bigr)
   \;=\; c^T \, a^2\, F_{\mu\nu}(0) \,+\, {\cal O}(a^3),
   \label{eq:70}
\end{equation}
where $\dov_{0,0}$ is the matrix element of the overlap operator at
$(m,n) = (0,0)$. The non-zero constant $c^T=c^T(\rho)$ is independent of
$A_\mu(x)$, and $\sigma_{\mu\nu}\equiv \frac{1}{2i}[\gamma_\mu,\gamma_\nu]$.}

\section{Calculation}

To proceed with the calculation, we shall assume that $\trs \smn \dov_{0,0}$ has a Taylor 
expansion in $a$ and we will compute the leading contributions.

To evaluate the diagonal element $\dov_{n,n}$ we introduce the momentum variable in the 
following way~\cite{ada02,suz99}
\beqan   \label{ft}
\dov_{n,n}&=&\sum_m \dov_{n,m} \delta_{n,m} = \sum_m \dov_{n,m} \intdk e^{-i k(n-m)} \\
&=&\intdk e^{-i k n} \sum_m \dov_{n,m} e^{i k m}.
\eeqan
This will allow us to evaluate the inverse square root in Eq.~(\ref{eq:50}).   
Next, we define the diagonal matrices $K(k)$ as
\beq
(K(k))_{n,m} \equiv e^{i k n} \delta_{n,m}.
\eeq
These matrices are unitary: $K^\dagger(k) = K^{-1}(k)=K(-k)$. If we now introduce the vector
$\one$ such that $\one_n = 1$, a vector with all entries set to $1$, we can rewrite the above
expression as
\beq      \label{ft2}
\dov_{n,n} = \intdk \left(K^{-1}(k) \dov K(k) \one\right)_n.
\eeq

To calculate
 $K^{-1}(k)\dov K(k)$, we assume that we can express $\frac{1}{\sqrt{X^\dagger X}}$ as
 a power series in $X^\dagger X$. Then
\beq   \label{pol}
 K^{-1}(k)\dov K(k) = \rho\lb \id-Y\frac{1}{\sqrt{Y^\dagger Y}} \rb,
 \eeq   
 where 
\beq
Y=-\bar{X}=-K^{-1}(k) X K(k) = M+\tR - \stD - i \ss,
\eeq
 and 
\beqan
 \tD_\mu &=& \half\lb e^{i k_\mu} (U_\mu S_\mu - 1) -
 e^{-i k_\mu}(S_\mu^\dagger U_\mu^\dagger -1)\rb, \nonumber \\
 \tR &=& \frac{r}{2}\sum_\mu \lb e^{i k_\mu} (U_\mu S_\mu - 1) +
 e^{-i k_\mu}(S_\mu^\dagger U_\mu^\dagger -1)\rb, \nonumber \\
 M &=& \rho + r\sum_\lambda(c_\lambda -1), 
\eeqan
where  $s_\mu = \sin k_\mu, \quad c_\mu = \cos k_\mu$.

 Eqs. (\ref{ft}), (\ref{ft2}), and (\ref{pol}),  lead to
 \beq
 \trs \smn \dov_{0,0} = -\rho \intdk \trs \smn \lb Y\frac{1}{\sqrt{Y^\dagger Y}}\one\rb_0.
 \eeq

 \subsection{Computational strategy}

 As we mentioned before, we assume that $\trs \smn \dov_{0,0}$ has a Taylor expansion
 in $a$. We will compute the leading contributions by taking derivatives with respect to $a$
 and then evaluating the limit $a\rightarrow 0$. We assume that all the matrices and
 matrix products are well defined and that we can take derivatives in the usual fashion. The
 non-trivial part of the calculation is taking the derivative of $\invsqYY$ with respect to $a$.
 For this we will write
 \beq
\invsqYY= \frac{1}{\pi} \intds \invsig{Y},
 \eeq
 under the condition that $Y^{\dagger}Y$ does not have zero eigenvalues, which is satisfied
for the case of classical gauge fields sufficiently close to the continuum limit~\cite{ov_loc}.
 Noting that $(M^{-1})' = - M^{-1} M' M^{-1}$  for a matrix $M$,
 we have
 \begin{eqnarray}   \label{derivative}
 \da \invsqYY &=& -\frac{1}{\pi} \intds \invsig{Y} \lb \da Y^\dagger Y\rb \invsig{Y} \nonumber \\
 \dasq \invsqYY &=& \frac{2}{\pi} \intds \invsig{Y} \lb \da Y^\dagger Y\rb \invsig{Y} 
\lb \da Y^\dagger Y\rb \invsig{Y}  \nonumber \\
 &&-\frac{1}{\pi} \intds \invsig{Y} \lb \dasq Y^\dagger Y\rb \invsig{Y}.
 \end{eqnarray}

We see that the problem is reduced to computing derivatives of $Y$ with respect to $a$. The only
matrices that depend on $a$ are the link matrices $U_\mu$. Since we are only interested in the
limit $a\rightarrow 0$ and since we will only calculate the contributions up to order $a^2$,
we only need the following limits
\beqan   \label{derU}
\lim_{a\rightarrow 0} U_\mu &=& \id, \nonumber \\
\lim_{a\rightarrow 0} \da U_\mu &=& i A_\mu(0) \id, \nonumber \\
\lim_{a\rightarrow 0} \dasq U_\mu &=& - A_\mu(0)^2 \id + 2 i N \nabla A_\mu(0).
\eeqan
where
\beq
(N\nabla A_\mu(0))_{m,n} \equiv \sum_\alpha n_\alpha \partial_\alpha A_\mu(0) \delta_{m,n},
\eeq
with $n_\alpha$ being the component of the 4-vector $n$.
 We see that in the limit $a\rightarrow 0$ both $U_\mu$ and $\da U_\mu$
reduce to an identity matrix in the space-time coordinates, but they are not necessarily
diagonal in color space. However, the second derivative $N \nabla A_\mu(0)$ has a term that is 
different: this matrix is still diagonal in the
space-time index (since it is the derivative of a diagonal matrix) but the diagonal elements 
are not equal.

To compute $(Y\invsqYY \one)_0$ we will need to justify several relations.

\noindent
$\bullet$
{\it Relation 1:}   \\
\beq    \label{Y0z}
Y_0^\dagger Y_0 \one = z \one,
\eeq
where 
\beqan
Y_0 &=&\lim_{a\rightarrow 0} Y = M+\tR_0+\stD_0+i\ss, \\
\tD_{0,\mu} &=& \lim_{a\rightarrow 0} \tD_{\mu} = \half\lb e^{i k_\mu} (S_\mu - 1) -
 e^{-i k_\mu}(S_\mu^\dagger -1)\rb, \nonumber \\
\tR_0 &=& \lim_{a\rightarrow 0} \tR = \frac{r}{2}\sum_\mu \lb e^{i k_\mu} (S_\mu - 1) +
 e^{-i k_\mu}(S_\mu^\dagger -1)\rb, 
\eeqan
and
\beq
z = \sum_\mu s_\mu^2+M^2,
\eeq
is a number.

To show Eq. (\ref{Y0z}), we write
\beqan
Y_0^\dagger Y_0 &=& \lb (M+\tR_0)+(\stD_0+i\ss) \rb \lb(M+\tR_0)-(\stD_0+i\ss)\rb \nonumber \\
&=& (M+\tR_0)^2 - (\stD_0+i\ss)^2 + [\stD_0 + i\ss, M+\tR_0].
\eeqan
The commutator
is zero since $[S_\mu, S_\nu]=0$ and thus we have
\beq
Y_0^\dagger Y_0 = M^2+\tR_0^2 + 2 M \tR_0 +\ss^2 - \stD_0^2 - i \{ \ss, \stD_0\}.
\eeq
It is easy to see that $\tR_0\one = 0$ and $\stD_0\one=0$ since the ``derivative" like term,
$S_\mu-1$, vanishes when acting on $\one$. This ``derivative" term comes from the continuum
limit of  $(U_\mu S_\mu-1)$ which becomes $(S_\mu-1)$ as $a\rightarrow 0$. We thus have 
\beq  
Y_0^\dagger Y_0 \one = (M^2 + \ss^2)\one = z \one.
\eeq

\noindent
$\bullet$
{\it Relation 2:}   \\
\beq    \label{Yz}
\lim_{a\rightarrow 0} \invsig{Y} \one = \invsig{Y_0}\one = \invsigz \one,
\eeq
  This can be straight-forwardly shown if we expand $\invsig{Y_0}$ as a power series in 
$Y_0^\dagger Y_0$ and apply {\it Relation 1} in Eq.~(\ref{Y0z}) successively.

\noindent
$\bullet$
{\it Relation 3:}   \\
\beq   \label{rel3}
\invsig{Y_0} N\nabla A_\mu(0)\one = \invsigz N\nabla A_\mu(0)\one -
\frac{1}{(\sigma^2+z)^2} \Delta N\nabla A_\mu(0)\one,
\eeq
where
\beq
\Delta = 2 M \tR_0 - i \{\ss,\stD_0\} = 2 M  \tR_0 - 2 i \sum_\mu s_\mu(\tD_\mu)_0. \label{delta}
\eeq
To compute the second order derivative in Eq.~(\ref{derivative}) we will need this relation  
when the matrix $\invsig{Y_0}$ acts on non-constant vectors of the form 
$N\nabla A_\mu(0)\one$ in Eq.~(\ref{derU}). To show this we write $Y_0^\dagger Y_0$ as
\beq
Y_0^\dagger Y_0 = z + \underbrace{\tR_0^2-\stD_0^2}_{\Delta_2} +
\underbrace{2 M \tR_0 - i\{\ss,\stD_0\}}_{\Delta},
\eeq
where $\Delta_2$ and $\Delta$ are unrelated. Using the fact that
\beqan   \label{derivative_NC}
(S_\nu-1) N\nabla A_\mu(0) \one &=& ((N+\hat{\nu})\nabla A_\mu(0)-N\nabla A_\mu(0))\one \nonumber \\
&=& \hat{\nu}\nabla A_\mu(0)\one = \partial_\nu A_\mu(0)\one,
\eeqan
which is a constant vector,  we can easily see that
$\Delta_2 N\nabla A_\mu(0)\one=0$ and $\Delta\Delta N\nabla A_\mu(0)\one=0$.
This is because these terms include double ``derivatives" like $(S_\mu-1)(S_\nu-1)$ and since
the first ``derivative" produces a constant vector, the second ``derivative" acting on it
makes it vanish. Now that we have
$Y_0^\dagger Y_0 N\nabla A_\mu \one = (z + \Delta) N\nabla A_\mu \one$
and $\Delta\Delta N\nabla A_\mu(0)\one=0$, we can prove by induction that
\beq   \label{induction}
(Y_0^\dagger Y_0)^k N\nabla A_\mu \one = z^k N\nabla A_\mu \one +
k z^{k-1} \Delta N\nabla A_\mu \one.
\eeq
From a series expansion $P(Y_0^\dagger Y_0)=\invsig{Y_0}$ and Eq.~(\ref{induction}),
we see that 
\beq \label{ind2}
P(Y_0^\dagger Y_0)N\nabla A_\mu \one = P(z)N\nabla A_\mu \one+
P'(z)\Delta N\nabla A_\mu \one.
\eeq 
Since $P(z)=\invsigz$ and
$P'(z)=-\frac{1}{(\sigma^2+z)^2}$, Eqs. (\ref{induction}) and (\ref{ind2}) lead
to {\it Relation 3} in Eq.~(\ref{rel3}).


As we mentioned before, we assume that our function admits a Taylor expansion. In this 
case, we can write
\beqan   \label{TE}
\trs \smn \lb Y\frac{1}{\sqrt{Y^\dagger Y}}\one\rb_0 &=&
\lim_{a\rightarrow 0} \trs \smn \lb Y \invsqYY \one \rb_0  \nonumber \\
&+& a \lim_{a\rightarrow 0} \trs \smn \da \lb Y \invsqYY \one \rb_0 \nonumber \\
&+& \half a^2 \lim_{a\rightarrow 0} \trs \smn \dasq \lb Y \invsqYY \one \rb_0 + O(a^3).
\eeqan
We will now proceed to carry out our calculation order by order.

\subsection{Calculation details}

\subsubsection{Order 0}

We need to compute
\beq
\lim_{a\rightarrow 0} \trs \smn \lb Y \invsqYY \one \rb_0 =
\frac{1}{\pi} \intds \trs \smn \lb Y_0\invsig{Y_0}\one \rb_0,
\eeq
where the relevant term is $\trs \smn Y_0\invsig{Y_0}\one=\invsigz\trs \smn Y_0\one$. Now
\beq          \label{Y0}
Y_0\one = (M + \tR_0 - i\ss - \stD_0)\one = (M-i\ss)\one,
\eeq
due to the fact $\tR_0\one=\stD_0\one=0$. Since $Y_0\one$ has only scalar and vector spinor
components and no tensor component,  it leads to $\trs \smn Y_0\one=0$ and the zeroth order 
contribution is thus zero, i.e:
\beq    \label{orer0}
\lim_{a\rightarrow 0} \trs \smn \lb Y \invsqYY \one \rb_0 = 0.
\label{order0}
\eeq

\subsubsection{Order 1}

We need to compute
\beqan
&&\lim_{a\rightarrow 0} \trs \smn \da \lb Y \invsqYY \one \rb_0 =
\frac{1}{\pi} \intds \trs \smn \lb Y'_0\invsig{Y_0}\one \rb_0 \nonumber \\
&& - \frac{1}{\pi} \intds \trs \smn \lb Y_0 \invsig{Y_0} (Y^\dagger Y)'_0\invsig{Y_0}\one\rb_0.
\eeqan
The first term contribution vanishes since
\beq
Y'_0\invsig{Y_0}\one = \invsigz Y'_0\one=\invsigz (\tR'_0-\stD'_0)\one,
\eeq
and then we are left again with only scalar and vector spinor components. 
Thus $\trs \smn Y'_0\invsig{Y_0}\one=0$.

To evaluate the second term we need to compute
$(Y^\dagger Y)'_0\one = (2 M \tR'_0 - 2 i \sum_\alpha s_\alpha \tD'_{\alpha 0})\one$ using
the following identities
\beqan
\tR'_0 &=& \frac{r}{2} \sum_\mu i A_\mu(0) \lb e^{ik_\mu} S_\mu - e^{-i k_\mu} S_\mu^\dagger \rb, \\
\tD'_{\mu 0} &=& \half i A_\mu(0) \lb e^{ik_\mu} S_\mu + e^{-i k_\mu} S_\mu^\dagger \rb.
\eeqan
We have then $\tR'_0\one = -r \sum_\alpha s_\alpha A_\alpha(0) \one$ and
$\tD'_{\alpha 0}\one = i c_\alpha A_\alpha(0)\one$ and
\beq
(Y^\dagger Y)'_0\one = 2 \sum_\alpha s_\alpha A_\alpha(0) \lb -M r + c_\alpha \rb \one,
\eeq
a constant vector. We have then
\beqan
&&Y_0 \invsig{Y_0} (Y^\dagger Y)'_0\invsig{Y_0}\one   \nonumber \\
&&=\invsigz Y_0 \invsig{Y_0} (Y^\dagger Y)'_0\one \nonumber \\
&&= \invsigz 2 \sum_\alpha s_\alpha A_\alpha(0) \lb -M r + c_\alpha \rb Y_0 \invsig{Y_0} \one \nonumber \\
&&= \lb \invsigz \rb^2 2 \sum_\alpha s_\alpha A_\alpha(0) \lb -M r + c_\alpha \rb Y_0\one \nonumber \\
&&= \lb \invsigz \rb^2 2 \sum_\alpha s_\alpha A_\alpha(0) \lb -M r + c_\alpha \rb ( M - i\ss)\one.
\eeqan
We see that we only have scalar and vector spinor components and this term vanishes too after
taking the spinor trace. Thus the first order contribution vanishes.
\beq   \label{order1}
\lim_{a\rightarrow 0} \trs \smn \da \lb Y \invsqYY \one \rb_0 = 0.
\label{order1}
\eeq

\subsubsection{Order 2}

   The derivation of the second order contribution is somewhat involved, but employs the
same steps as above. The details of the calculations are presented in \mbox{Appendix A} for the 
perusal of interested readers. The main result is that
the second order contribution is not zero and is given by
\beqan   
&&\intdk \lim_{a\rightarrow 0} \trs \smn \dasq \lb Y \invsqYY \one \rb_0 \nonumber \\
&&= - \intdk \frac{4 (M c_\mu c_\nu + r s_\mu^2 c_\nu + r s_\nu^2 c_\mu)}{z^{3/2}} F_{\mu\nu}(0). 
\label{order2}
\eeqan
Together with the (null) results from the zeroth and first orders in \mbox{Eqs.~(\ref{order0}) and
~(\ref{order1}),} the final result is

\beq  
\trs \smn \dov_{0,0} = a^2 F_{\mu\nu}(0)
\rho \intdk \frac{2 (M c_\mu c_\nu + r s_\mu^2 c_\nu + r s_\nu^2 c_\mu)}{z^{3/2}} + O(a^3).
\label{FR}
\eeq

    Comparing with Eq.~(\ref{ct}), we find

\beq
c^T(\rho)= \rho \intdk \frac{2 (M c_\mu c_\nu + r s_\mu^2 c_\nu + r s_\nu^2 c_\mu)}{z^{3/2}}.
\eeq
With $r=1$ and $\rho= 1.368$ (which corresponds to $\kappa= 0.19$ in the Wilson Dirac
operator), we find $c^T= 0.11157$.~\footnote{We should remark that $D_{\mu}$ and $R$ in 
Eq.~(\ref{DR}) can be written as
$D_\mu (x)= \half [U_\mu (x) e^{a\partial_{\mu}} - e^{-a\partial_{\mu}} U_\mu^\dagger(x)]$ 
and
$R(x) = \frac{r}{2} \sum_\mu [U_\mu(x) e^{a\partial_{\mu}}+  e^{-a\partial_{\mu}}U_\mu^\dagger(x)]$,
such as defined in Ref.~\cite{suz99}. Upon taking derivatives with respect to
$a$ in $e^{\pm a\partial_{\mu}}$ and $U_{\mu}$ in Eq.~(\ref{TE}), the results
in Eqs.~(\ref{order0}), (\ref{order1}), (\ref{order2}), and (\ref{FR}) were obtained~\cite{liu06}.
However, due to the fact that $\sigma_{\mu\nu}$ in Ref.~\cite{liu06} is defined with an opposite 
sign from the one used here which is $\sigma_{\mu\nu}\equiv \frac{1}{2i}[\gamma_\mu,\gamma_\nu]$, 
the result in Ref.~\cite{liu06} is negative of that in Eq.~(\ref{FR}).}

\section{Conclusions}

 We have shown in an explicit calculation that, for the overlap Dirac operator, 
the classical continuum limit of 
$\trs \smn D^{ov}(x,x)$ is proportional to the gauge field strength tensor $F_{\mu\nu}(x)$.

   Based on the experience of studying the QCD vacuum structure with
   the topological charge density defined from the overlap operator,
   it is found that one can obtain clear signals with only a handful
   of gauge
   configurations~\cite{hdd03b,haz05a,ahz05,ikk06,alt05,haz05b}. This
   is presumably due to the non-ultralocal nature of the overlap
   operator which serves as an efficient filter of the ultraviolet
   fluctuations~\cite{hov06a,hov06b}. It is worthwhile then to study
   whether other operators defined in a similar fashion share this
   property. For example, it would be interesting to see if the
   calculation of glueball masses, glue momentum and angular momentum
   in the nucleon, etc. can benefit from employing the overlap-based
   definition of the field strength tensor
%
\begin{equation}
      {\cal O}(x) \,\equiv \, \frac{1}{c^T} \, 
      {\Trs}_s \, \sigma_{\mu\nu} D^{ov}(x,x)  
\end{equation}
which is properly normalized. We should point out
that the value of the constant $c^T=c^T(\rho)$ depends on the mass parameter $\rho$ used
to define the overlap operator. We will study this $\rho$--dependence 
in detail elsewhere~\cite{ahl07}. 

Finally, we wish to mention that, for the purposes of studying QCD vacuum 
structure it is useful to be able to expand gauge observables in low--lying Dirac eigenmodes.
Indeed, such expansions in the case of overlap--based topological density proved
to be useful in studying the low--energy behavior of the topological vacuum
structure~\cite{hdd02,haz05a}. Thus, one rationale for defining all
gauge operators in terms of Dirac kernels is the fact that it allows such
an expansion for an arbitrary operator~\cite{hov06b}. We note that for the 
purpose of eigenmode expansion, the expression for the field strength tensor 
in terms of the squared lattice Dirac
operators was also considered in Ref.~\cite{gat02}.

\newpage

\begin{center}
\appendix
{{\Large Appendix A}}
\end{center}

We need to compute
\beqan    \label{a2}
\lim_{a\rightarrow 0}\! &\trs&\!\!\!\! \smn \dasq \lb Y \invsqYY \one \rb_0 =
\frac{1}{\pi} \intds \trs \smn \lb Y''_0\invsig{Y_0}\one \rb_0   \nonumber \\
&-&\frac{2}{\pi} \intds \trs \smn \lb Y'_0 \invsig{Y_0} (Y^\dagger Y)'_0\invsig{Y_0}\one\rb_0
\label{a2term2}  \nonumber \\
&+&\frac{2}{\pi} \intds \trs \smn \lb Y_0 \left[\invsig{Y_0} (Y^\dagger Y)'_0 \right]^2\invsig{Y_0}\one\rb_0 \label{a2term3} \nonumber \\
&-& \frac{1}{\pi} \intds \trs \smn \lb Y_0 \invsig{Y_0} (Y^\dagger Y)''_0\invsig{Y_0}\one\rb_0 \label{a2term4}
\eeqan
The first term in  Eq. (\ref{a2}) turns out to be zero, since the expression
\beq
Y''_0\invsig{Y_0}\one = \invsigz Y''_0\one = \invsigz (\tR''_0 - \stD''_0)\one,
\eeq
only has scalar and vector spinor components and, as a result, it vanishes upon taking the 
the spinor trace in Eq. (\ref{a2}). The
second term in  Eq. (\ref{a2}) also vanishes, because
\beq
Y'_0 \invsig{Y_0} (Y^\dagger Y)'_0\invsig{Y_0}\one =
\lb \invsigz \rb^2 2 \sum_\alpha s_\alpha A_\alpha(0) \lb -M r + c_\alpha \rb Y'_0\one
\eeq
also has only scalar and vector components. The third term in  Eq. (\ref{a2}) vanishes for the
same reason, since
\beq
Y_0 \left[\invsig{Y_0} (Y^\dagger Y)'_0 \right]^2\invsig{Y_0}\one =
\lb \invsigz \rb^3 \lb 2 \sum_\alpha s_\alpha A_\alpha(0) \lb -M r + c_\alpha \rb\rb^2 Y_0\one
\eeq
has only spinor and vector components.

The only non-zero contributions come from the fourth term in  Eq. (\ref{a2}). To compute its
contribution, we need to evaluate
\beq  \label{YY2}
(Y^\dagger Y)''_0 \one=(Y_0^\dagger Y''_0+{Y''_0}^\dagger Y_0+2 {Y'_0}^\dagger Y'_0)\one.
\eeq
We will compute each term separately. For the first term, we have
\beq
Y''_0 = \tR''_0 - \stD''_0 = -\half \sum_\mu \lb e^{i k_\mu} (\gamma_\mu-r)
 U''_{\mu 0} S_\mu - e^{-i k_\mu} (\gamma_\mu + r) S_\mu^\dagger
{U''_{\mu 0}}^\dagger \rb, \label{y0dd}
\eeq
where $U''_{\mu 0} = -A_\mu(0)^2+2iN\nabla A_\mu(0)$. In this case, $Y''_0\one$ will have a constant
vector part
\beq
(Y''_0\one)_c = \sum_\mu  \lb A_\mu(0)^2(i s_\mu\gamma_\mu-r c_\mu)+ie^{-i k_\mu}(\gamma_\mu + r)
\partial_\mu A_\mu(0) \rb \one,
\eeq
and a non-constant part
\beq
(Y''_0\one)_{nc} = -2 i \sum_\mu N \nabla A_\mu(0) (c_\mu\gamma_\mu-i r s_\mu)\one.
\eeq
In considering the first term in Eq. (\ref{YY2}), we note that the derivative terms in 
$Y_0^\dagger$, i.e. $\tR_0^{\dagger}+\stD_0^{\dagger}$, vanish when acting on constant
vectors as shown in Eq. (\ref{Y0}). Furthermore, the derivatives acting on the non-constant part 
produce constant vectors as in Eq.~(\ref{derivative_NC}). As a result, we get a constant term and a non-constant
term
\beqan  \label{y0y0dd}
(Y_0^\dagger Y''_0\one)_c &=& (M+i\ss)\sum_\mu \left[ A_\mu(0)^2(i s_\mu \gamma_\mu- r c_\mu)+
i e^{-i k_\mu}(\gamma_\mu + r)\partial_\mu A_\mu \right] \one \nonumber \\
&-& 2 i \sum_{\mu,\nu} (c_\mu\gamma_\mu + i r s_\mu)
(c_\nu\gamma_\nu - i r s_\nu)\partial_\mu A_\nu(0)\one, \nonumber \\
(Y_0^\dagger Y''_0\one)_{nc} &=& -2 i (M+i\ss)\sum_\mu N\nabla A_\mu(0)(c_\mu\gamma_\mu-i r s_\mu)
\one. 
\eeqan

From Eq.~(\ref{y0dd}), we have
\beq
{Y''_0}^\dagger = \tR''_0 + \stD''_0 = \half \sum_\mu \lb e^{i k_\mu} (\gamma_\mu+r)
 U''_{\mu 0} S_\mu - e^{-i k_\mu} (\gamma_\mu - r) S_\mu^\dagger
{U''_{\mu 0}}^\dagger \rb.
\eeq
Consequently, the second term, ${Y''_0}^\dagger Y_0\one$, gives
\beqan   \label{y0ddy0}
({Y''_0}^\dagger Y_0\one)_c &=& -\sum_\mu\left[ A_\mu(0)^2(i s_\mu \gamma_\mu + r c_\mu) +
i e^{-i k_\mu}(\gamma_\mu - r)\partial_\mu A_\mu(0)\right](M- i\ss)\one,  \nonumber \\
({Y''_0}^\dagger Y_0\one)_{nc} &=& 2 i \sum_\mu N\nabla A_\mu(0) (c_\mu \gamma_\mu + i r s_\mu)
(M-i\ss)\one. 
\eeqan

The last term to evaluate is $2 {Y'_0}^\dagger Y'_0\one$. We note that since this term only involves
first derivatives it will only produce a constant vector. Using 
\beq
Y'_0 = -\half\sum_\mu i A_\mu(0)\lb e^{i k_\mu}(\gamma_\mu-r)S_\mu + e^{-i k_\mu}(\gamma_\mu+r)
S_\mu^\dagger \rb,
\eeq
we get
\beq        \label{y0dy0d}
2 {Y'_0}^\dagger Y'_0\one = 2 \sum_{\mu,\nu} A_\mu(0) A_\nu(0) (c_\mu\gamma_\mu+ i r s_\mu)
(c_\nu\gamma_\nu-i r s_\nu)\one. 
\eeq

Putting all the contributions from Eqs.~(\ref{y0y0dd}), (\ref{y0ddy0}), 
and (\ref{y0dy0d}) together, we obtain 
\beqan    \label{cterm}
((Y^\dagger Y)''_0\one)_c &=& \sum_\mu \left[ A_\mu(0)^2(-2 M r c_\mu - 2 s_\mu^2) +
i e^{-i k_\mu} \partial_\mu A_\mu(0)(2 M r + 2 i s_\mu)\right] \one \nonumber \\
&+&  2 \sum_{\mu,\nu} (A_\mu(0) A_\nu(0)-i\partial_\mu A_\nu(0)) (c_\mu\gamma_\mu+ i r s_\mu)
(c_\nu\gamma_\nu-i r s_\nu)\one,
\eeqan
and
\beq     \label{ncterm}
((Y^\dagger Y)''_0\one)_{nc} = 4 \sum_\mu N\nabla A_\mu(0) s_\mu (c_\mu - M r)\one. 
\eeq

We now return to evaluating the last term of the second order contribution in  Eq.~(\ref{a2})
\beqan
&&Y_0 \invsig{Y_0} (Y^\dagger Y)''_0\invsig{Y_0}\one =
\invsigz Y_0 \invsig{Y_0} (Y^\dagger Y)''_0 \one  \nonumber \\
&&= \lb \invsigz \rb^2  Y_0 ((Y^\dagger Y)''_0\one)_c +
 \invsigz Y_0 \invsig{Y_0} ((Y^\dagger Y)''_0 \one)_{nc}
\eeqan
From Eq.~(\ref{rel3}), we find for the non-constant term contribution
\beq   \label{nc_contri}
Y_0 \invsig{Y_0} ((Y^\dagger Y)''_0 \one)_{nc} = \invsigz Y_0 ((Y^\dagger Y)''_0 \one)_{nc}
- \lb \invsigz \rb^2 Y_0 \Delta ((Y^\dagger Y)''_0 \one)_{nc}.
\eeq
It is clear from Eq.~(\ref{ncterm}) that the non-constant term $((Y^\dagger Y)''_0 \one)_{nc}$ is a scalar.
Furthermore, from Eq.~(\ref{delta}), we see that $\Delta$ is a scalar too.
Since $Y_0$  has only scalar and vector components, the terms in Eq.~(\ref{nc_contri}) above 
have the same spinor structure.  Thus, after taking the spinor trace, all these 
terms vanish.

Putting together the above results, we have
\beq
\lim_{a\rightarrow 0} \trs \smn \dasq \lb Y \invsqYY \one \rb_0 =
- \frac{1}{\pi} \intds \lb\invsigz\rb^2 \trs \smn \lb Y_0 ((Y^\dagger Y)''_0\one)_c\rb_0.
\eeq
One can perform the integration over $\sigma$, since $Y_0 ((Y^\dagger Y)''_0\one)_c$
has no $\sigma$ dependence. We then have 
\beq
\lim_{a\rightarrow 0} \trs \smn \dasq \lb Y \invsqYY \one \rb_0 = -\frac{1}{2 z^{3/2}}
 \trs \smn \lb Y_0 ((Y^\dagger Y)''_0\one)_c\rb_0.
\eeq
Since $((Y^\dagger Y)''_0\one)_c$ is a constant vector, when $Y_0$ acts from the left, the
derivative terms in $Y_0$ vanish and, as a result, 
we have $Y_0((Y^\dagger Y)''_0\one)_c=(M-i\ss)((Y^\dagger Y)''_0\one)_c$. 
Since the first term in Eq.~(\ref{cterm}) is a scalar, its contribution vanishes after taking the trace. 
We have then
\beqan
&&\trs \smn Y_0 ((Y^\dagger Y)''_0\one)_c   \nonumber \\
&&=2 \trs \smn (M-i\ss) \sum_{\alpha,\beta}
 (A_\alpha(0) A_\beta(0)-i\partial_\alpha A_\beta(0)) (c_\alpha\gamma_\alpha+ i r s_\alpha)
 (c_\beta\gamma_\beta-i r s_\beta)\one  \nonumber \\
&& =2 \trs \smn \sum_{\alpha,\beta}
 (A_\alpha(0) A_\beta(0)-i\partial_\alpha A_\beta(0))
 (M c_\alpha c_\beta \gamma_\alpha \gamma_\beta + r s_\alpha c_\beta \ss \gamma_\beta
 - r s_\beta c_\alpha \ss \gamma_\alpha)\one. \nonumber \\
&&                                                    
\eeqan
To finish our calculation, we will use the fact that $z$ and $M$ are even functions of $k_\mu$
and thus any integral over $k_\mu$ that involves an odd power of $s_\mu$ will vanish. For the spinor
trace we use the relation $\trs \smn \gamma_\alpha\gamma_\beta = 4 i
(\delta_{\mu\alpha}\delta_{\nu\beta} - \delta_{\mu\beta}\delta_{\nu\alpha})$ and we finally obtain
\beqan   
&&\intdk \lim_{a\rightarrow 0} \trs \smn \dasq \lb Y \invsqYY \one \rb_0 \nonumber \\
&&= - \intdk \frac{4 (M c_\mu c_\nu + r s_\mu^2 c_\nu + r s_\nu^2 c_\mu)}{z^{3/2}} F_{\mu\nu}(0). 
\eeqan


\begin{center}
{\large Acknowledgment}
\end{center}

   This work is partially supported by the U.S. DOE grant DE-FG05-84ER40154.

\end{document}